\RequirePackage[2020-02-02]{latexrelease}
\documentclass[aps,pre,twocolumn,showpacs,superscriptaddress,groupedaddress]{revtex4-2}
\usepackage{amsmath}
\usepackage{amssymb}
\usepackage{amsfonts}
\pdfoutput= 1
\usepackage{graphicx}
\usepackage{dcolumn}
\usepackage{bm}
\usepackage{sidecap}
\usepackage{subfloat}
\usepackage{parskip}
\usepackage{grffile}
\usepackage{color}
\usepackage{hyperref}
\usepackage{hhline}
\usepackage{mathtools}
\usepackage{graphics}
\usepackage{multirow}
\usepackage{verbatim}
\usepackage{longtable}
\usepackage{rotating}
\usepackage{setspace}
\usepackage{epsfig}
\usepackage{subfigure}
\usepackage{epstopdf}
\usepackage{gensymb}
\usepackage[normalem]{ulem}
\usepackage{tikz}

\makeatletter
\def\@eqnnum{{\normalsize \normalcolor (\theequation)}}
\makeatother

\graphicspath{{./}{ER/main/}}

\begin{document}

\title{Eigenvector localization in hypergraphs: pair-wise vs higher-order links}
\author{Ankit Mishra} \email{ankitphy0592@gmail.com}
\author{Sarika Jalan} \email{sarika@iiti.ac.in}
\affiliation{Complex Systems Lab, Department of Physics, Indian Institute of Technology Indore, Khandwa Road, Simrol, Indore-453552}

\begin{abstract}

Localization behaviours of Laplacian eigenvectors of complex networks provide understanding to various dynamical phenomena on the corresponding complex systems.
We numerically investigate role of hyperedges in driving eigenvector localization of hypergraphs Laplacians. By defining a single parameter $\gamma$ which measures the relative strengths of pair-wise and higher-order interactions, we analyze the impact of interactions on localization properties. For $\gamma \leq 1$, there exists no impact of pair-wise links on eigenvector localization while the higher-order interactions instigate localization in the larger eigenvalues. 
For $\gamma >1$, pair-wise interactions cause localization of eigenvector corresponding to small eigenvalues, where as higher-order interactions, despite being much lesser than the pair-wise links, keep driving localization of the eigenvectors corresponding to larger eigenvalues.
The results will be useful to understand dynamical phenomena such as diffusion, and random walks  on a range of real-world complex systems having higher-order interactions.

\end{abstract}

\maketitle

\section{Introduction}
Network science has emerged as a powerful tool to describe many real-world complex systems \cite{Rev_network,*SN_Dor,*newman_siam}. Networks consist of nodes, which are system's constituent elements and links represent the interactions among the nodes. Many real-world systems inherently share common structural properties, such as power-law degree distribution, short path-length, high clustering, degree-degree correlation, etc. These structural properties govern various dynamical processes on networks, for instance, disease spreading \cite{dis_struc_1, *dis_struc_2,*dis_struc_3}, stead-state behaviours of random walkers \cite{RW_struc}, synchronization \cite{arenas_synch}, etc. Despite tremendous theoretical success of network theory for real-world systems, it often becomes insufficient to capture properties and origin behind dynamical behaviors of many real-world systems accurately. With the increasing accumulation of data, it has been realized that often real-world interactions occur among more than two nodes at a time, while the network theory is best described for the binary relationship between nodes \cite{PV_HO}. Thus, the inefficacy of networks to model many-body interactions  compelled researchers to look beyond the realm of pair-wise interactions and develop appropriate higher-order models. The two most popular approaches for modeling higher-order interactions are hypergraphs \cite{HG_1,*HG_2,*HG_3} and simplicial complexes\cite{Simplex_1,*Simplex_2,*Simplex_3}.

Hypergraphs capturing higher-order interactions provide a more generalized model for real-world complex systems. A hypergraph consists of nodes and hyperedges; a hyperedge connects $d$ nodes at a time where typically $d \geq 2$. The size of the hyperedges of a hypergraph may differ, but if all the hyperedges consist of the same number of nodes $d$, it is referred to as $d-$uniform hypergraph. Thus, a $2-$uniform hypergraph would correspond to conventional graphs. Hypergraphs have been successfully used to model various real-world interactions such as biological \cite{Hypg_biology_1,*Hypg_biology_2}, social \cite{HG_social_1,*HG_social_2}, evolutionary dynamics \cite{HG_Evo_dyn_1,*HG_Evo_dyn_2}, etc. A simplicial complex is a special type of hypergraph which is closed under the subset operation. A simplicial complex not only consists of nodes and links but also incorporates structures of higher-order dimensions, like triangles, tetrahedrons, etc. Accordingly, a $k-$simplex describes simultaneous interactions between $k+1$ nodes forming a set $I = \{v_{1}, v_{2},v_{3},\dots v_{k+1}$\}. Thus, $1-$simplex corresponds to pair-wise interactions with  $I = \{v_{1}, v_{2}$\}, $2-$simplex would mean triangles with  $I = \{v_{1}, v_{2},v_{3}$\} and so on. 
A fundamental difference between modeling many-body interactions with the hypergraphs and simplicial complexes is that a $k-$simplex subsumes all the possible $k-1$ dimensions simplices which is not true with the former.

The spectra of adjacency and Laplacian matrices are known to affect various dynamical processes on networks. For example, epidemic threshold of infection rate on graphs is determined by the inverse of the principal eigenvalue of the corresponding adjacency matrices \cite{dis_thresh}. Spectral dimension provides a tool to characterize or determine the late-time behavior of diffusion \cite{spec_dimen_1, *spec_dimen_2}. Further, under appropriate conditions, the critical coupling strength at which synchronization occurs in coupled oscillators on networks is determined by the largest eigenvalue of the adjacency matrix \cite{onset_synch}. Additionally, spectra of the adjacency matrix also give insight into the structural properties of corresponding complex networks \cite{alok_2, Sj_1}. Likewise, spectra of the Laplacian matrices are related to the diffusion and other spreading phenomena on networks \cite{Lapl_flow}. The ratio of the largest to the first non-zero eigenvalue of a Laplacian matrix helps in determining the stability of generalized synchronization in the coupled dynamical system \cite{ratio_synch_1,*ratio_synch_2}. 
Further, Refs.~\cite{topo_synch_1,*topo_synch_2,*topo_synch_3}, highlight the existence of relationships among topological, spectral, and dynamical properties of networks. In addition to the eigenvalues, eigenvectors of the underlying adjacency and the Laplacian matrices have also been shown to be useful in providing insight into various structural and dynamical properties of the corresponding systems. In particular, localization of eigenvectors plays a crucial role in disease spreading \cite{PEV_disease}, perturbation propagation in ecological networks, \cite{PEV_ecological} etc. Further, localization of the Laplacian eigenvectors is useful in characterizing or identifying community structures \cite{APL_RMT_1,*APL_RMT_2}, stability of system against external shocks \cite{Econ_1}, network-turing patterns, \cite{N_turing_1,*N_turing_2} etc.

Though there exists a huge amount of literature on spectra of networks and, more specifically, eigenvector localization and their implications, no attempt has been made to understand  localization properties of eigenvectors for the higher-order interactions networks. Here, we consider hypergraphs as a higher-order model to probe the localization properties of eigenvectors. We study hypergraphs instead of simplicial complexes as the latter involves complicated combinatorics, and thus the applications are limited to the lower-order dimensions simplexes such as triangles and tetrahedrons \cite{HG_RW_1}. On the other hand, in the case of the hypergraphs, the information of higher-order structures is installed in a matrix form with the dimension equal to the number of nodes. Further, hypergraphs allow handling heterogeneous sizes of the hyperedges more efficiently than the simplicial complexes. Recent work on hypergraphs includes random walks \cite{HG_RW_1,HG_RW_2}, synchronization \cite{HG_Synch_1,HG_Synch_2,HG_Synch_3}, social contagion \cite{HG_dis_1,HG_dis_2}, evolutionary dynamics \cite{HG_Evo_dyn_1,HG_Evo_dyn_2}, etc. Most of these works, revolving around hypergraphs, focus on projecting hypergraph into its weighted pair-wise network and thereupon compare structural and dynamical properties between the hypergraphs and the projected pair-wise networks.
In this work, we take a slightly different approach, and instead of projecting a hypergraph into a corresponding pair-wise network, we consider contributions from the higher-order and pair-wise interactions for each node, and compare relative contributions of these two types of interactions in steering localization of the eigenvectors of hypergraph Laplacians.
We show that eigenvectors are localized on those nodes which have their pair-wise or higher-order degrees considerably deviating from the average higher-order degree ($\langle k^{p} \rangle$) and average pair-wise degree ($\langle k^{h} \rangle$) of the hypergraph, respectively.

The paper is organized as follows. Sec.~\ref{a1} consists of definitions of the Laplacian matrices of hypergraphs. Sec.~\ref{a2} introduces small-world hypergraphs and their basic properties. Sec. ~\ref{a3} discusses the methodology and techniques involved in the paper.
Sec. ~\ref{a4} contain results about impacts of the interplay of pair-wise and higher-order links on eigenvector localization.  Finally, Sec.~\ref{a5} concludes the paper with future directions.

\section{Laplacian Matrix} \label{a1}
A hypergraph denoted by ${\it H} = \{V,\,E^{H}\}$ consists of set of {\it nodes} and  {\it hyperedges}.\ The set of {\it nodes} are represented by $V$ = $\{v_{1}, v_{2}, v_{3}, $\ldots$, v_{N}\}$ and 
{\it hyperedges}
by $E^{H}$ = $\{E_{1}, E_{2}, E_{3}, $\ldots$ ,E_{M}\}$ where  $N$  and  $M$  are size of $V$ and $E^{H}$ respectively.\ Note that, each hyperedge $E_{\alpha}$, $\forall \alpha = 1,2,$\ldots$ ,M$, will contain a collection of nodes i.e.  $E_{\alpha} \subset V$. Thus, when $|E_{\alpha}| = 2$ for all $\alpha$, the hypergraph reduces to standard graph.
Mathematically, a hypergraph can be represented by its incidence matrix $(e_{i\alpha})_{N \times M}$ whose elements 
are defined as
\begin{equation}
\label{eq:incid}
e_{i \alpha}=\begin{cases} 1 &\text{$v_{i}\in E_{\alpha}$}\\
0 & \text{otherwise}\, .
\end{cases}
\end{equation}
One can easily construct the $N \times N$ adjacency matrix for a hypergraph using Eqn.~\ref{eq:incid}, as $A = ee^{T}$. The entries of the adjacency matrix $A_{ij}$
represents the number of hyperedges containing both the nodes $i$ and $j$. It is important to note here that the adjacency matrix is often obtained by setting $0$ to the main diagonal. We  further define $M \times M$ hyperedges matrix $C = e^{T}e$, whose entries $C_{\alpha \beta}$ represent the number of nodes  common between the hyperedges $E_{\alpha}$ and $E_{\beta}$.

There is no unique way to define the Laplacian matrix, $L$, of a hypergraph \cite{Phy_beyond}. One of the conventional way is as follows; $L_{ij} =k_{i} \delta_{ij} - A_{ij}$ where $k_{i} = \sum_{j =1}^{N} {A_{ij}}$  denotes the number of hyperedges containing the node $i$. However, it is not consistent with the full higher-order structures encrypted in the hypergraph. More specifically, it does not account for the sizes of the hyperedges incident on a node. Ref.~\cite{HG_RW_1}
solved this limitation by defining a new Laplacian matrix for a random walk which is also consistent with the higher-order structures. The transition probability of the random walker defined in \cite{HG_RW_1} takes care of the size of the hyperedges involved. More precisely, the Laplacian of the random walk defined in \cite{HG_RW_1} is as follows
\begin{equation} \label{eq:LRW}
L_{ij}^{RW}    =  \delta_{ij}-\frac{k^H_{ij}}{\sum_{i \neq l} k^H_{i\ell}},
\end{equation}
and the entries of $K^{H}$ matrix are given by
\begin{equation}
\label{eq:KH}
k^{H}_{ij}=\sum_{\alpha}(C_{\alpha \alpha}-1)e_{i\alpha}e_{j\alpha}=(e\hat{C}e^T)_{ij}-A_{ij}\quad\forall i\neq j\, 
\end{equation}
where ${\hat C}$ is a matrix whose diagonal entries coincide with that of $C$ and other entries are $zero$. Using Eqs.~\ref{eq:LRW} and ~\ref{eq:KH} \cite{HG_RW_2}, we construct the combinatorial Laplacian matrix for the hypergraph, given by
\begin{equation} \label{eq:Lapl}
L^{H} = K^{H} - D    
\end{equation}
Here, $D$ is the diagonal matrix  whose entries are  $D_{ii} =  k^{H}_{i} = {\sum_{i \neq \ell} k^H_{i\ell}}$, and $zero$ otherwise. Note that, in accordance with the earlier convention i.e. setting $0$ to the main diagonal, $k_{ii} = 0$. The Laplacian matrix  is given by Eqn.~\ref{eq:Lapl} takes into account both the number and size of the hyperedges incident on the nodes, and thus incorporates the higher-order structures completely. By considering $L^{H}$ as a Laplacian of the hypergraph, here we study the effect of higher-order structures on steering the eigenvector localization.
\begin{figure}[t]
	\centering
	\includegraphics[width= 0.4\textwidth ]{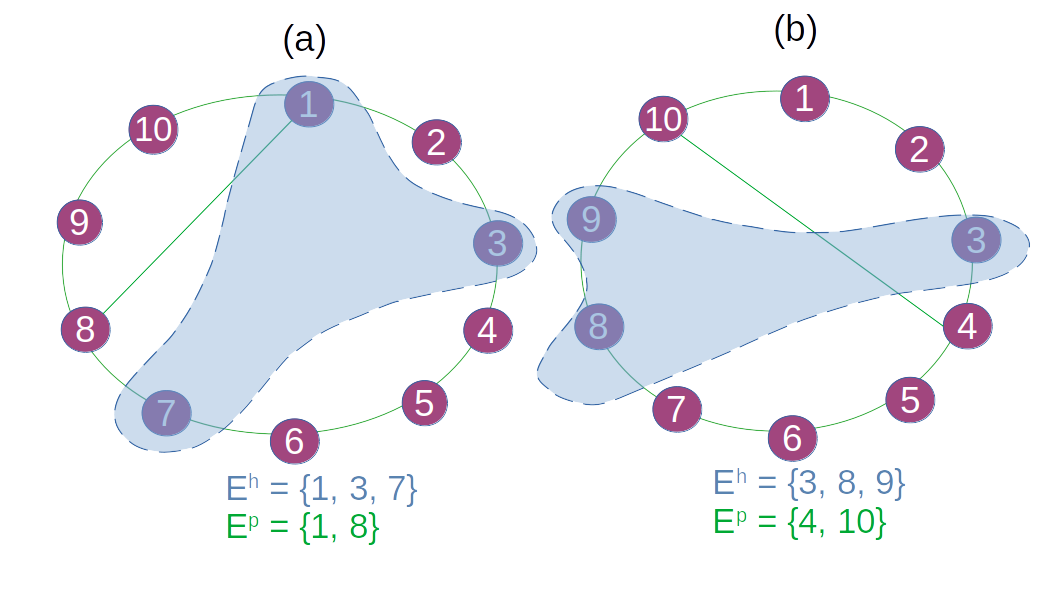} 
	\caption{(Color online) Schematic diagram of hypergraph model used here with $N = 10$, $M^{h} = 1$ and $M^{p} = 1$ for two different realizations. (a) $One$ pair-wise link, $E^{p} = \{1,8\}$ and $one$ hyperedge, $E^{h} = \{1,3,7\}$ are added into the ring lattice. (b)  $One$ pair-wise link, $E^{p} = \{4,10\}$ and $one$ hyperedge, $E^{h} = \{3,8,9\}$ are added into the ring lattice. The pair-wise links are solid lines (green) and hyperedge with dashed line (sky-blue) enclosing the involved nodes.}
	\label{SWH_model}
\end{figure}

\section{Model} \label{a2}
There are various ways in which a random hypergraph can be constructed  \cite{Phy_beyond}. We generate the hypergraph in this work as follows. First, a ring lattice is constructed in which each node is connected to its nearest neighbors on both the sides. We then randomly choose $d$ nodes uniformly from all the existing nodes. If already there is no hyperedge comprising of the chosen $d$ nodes, we add a hyperedge consisting of these $d$ nodes. 
For the simplicity, we consider $d = 3$, for each iteration. Next, we add pair-wise links by choosing $d = 2$ nodes uniformly and randomly from the existing nodes. The pair-wise links are added to the model so that an interplay of the higher-order and pair-wise links on the eigenvector localization can be investigated. The schematic diagram of the model is illustrated  in Fig.~\ref{SWH_model}. Note that, a similar model was also used in \cite{SWH_spin}, but no pair-wise links were added to the original ring lattice. We would like to further mention here that one can choose an alternate algorithm to generate the given model as introduced in \cite{Newman_SW}. In the alternate approach, for each node, one can choose $two$ other nodes with a probability $p$, and add hyperedges containing the nodes under consideration. Similarly, one can add pair-wise links by associating $one$ node for a given node. Thus, the total number of hyperedges, and pair-wise links, each, will be  equal to $pN$. However, in one loop  only $N$ pair-wise links can be added for $p = 1$.
To add more pair-wise links one has to again repeat the entire algorithm. Hence, we use the former algorithm in which  the number of pair-wise links and hyperedges are 
known from the beginning.

\section{Methods} \label{a3}
As discussed earlier, a hypergraph can be represented by its Laplacian matrix
$L^{H}$. Let the eigenvalues of the Laplacian matrix denoted by $\left\{\lambda_{1}, \lambda_{2}, \lambda_{3},\ldots,\lambda_{N}\right\}$ where
$\lambda_{1} \geq \lambda_{2} \geq $\ldots$ \geq \lambda_{N}$ and 
the corresponding orthonormal eigenvectors
as $\left\{\bm{x}_{1}, \bm{x}_{2}, \bm{x}_{3}, \ldots, \bm{x}_{N}\right\}$.
The Laplacian matrix is positive semi-definite, i.e., ${\sum_{i,j} L^H_{i,j}x_{i}x_{j}} \geq 0$ for any vector $\bm x = ({x}_{1}, {x}_{2}, {x}_{3}, \ldots, {x}_{N})$. Therefore, all the eigenvalues of the Laplacian matrix are positive with one and only one being zero for the connected network. The entries of eigenvector corresponding to this $zero$ eigenvalue will be uniformly distributed $({1,1, \dots ,1})/{\sqrt{N}}$. The generalized degree of a node $i$ in the hypergraph is given by $k^{H}_{i}$ which can be further decomposed into $k^{H}_{i} = k^{h}_{i}+k^{p}_{i}$ where $k^{h}_{i}$ and $k^{p}_{i}$ are the contributions from the higher-order and the pair-wise links, respectively. 
Similarly, the average degree, $\langle k \rangle = \frac{\sum_{i}{k^{H}_{i}}}{N}$, can be decomposed as $\langle k \rangle = \langle k \rangle^{h} + \langle k \rangle^{p}$ where
$\langle k \rangle^{h} = \frac{\sum_{i}{k^{h}_{i}}}{N} $ and $\langle k \rangle^{p} = \frac{\sum_{i}{k^{p}_{i}}}{N}$. 
We would like to further define $\langle k \rangle $ in terms of total number of the higher-order and pair-wise links as $\langle k \rangle = \frac{2 \times M^{p}}{N}+ \frac{12 \times M^{h}}{N}$, where $M^{p}$ and $M^{h}$ are total number of the pair-wise edges ($d = 2$) and the hyperedges ($d= 3$) in the hypergraph.
Also, it is important to note that if a node, say $i$, gets $1$ additional pair-wise links and $1$ higher-order links, its degree will be increased by $1$ and $4$ from the pair-wise and higher-order links, respectively. For example, if we consider the hypergraph depicted in Fig.~\ref{SWH_model}[a], the first row of $K^{H}$ matrix ($i = 1$) is the following,  $K^{H}_{1j} = [0,1,2,0,0,0,2,1,0,1]$. Notice that, $K^{H}_{13} = 2,K^{H}_{17} = 2$ from the hyperedge. Therefore, $k^{H}_{i} = 7$ with $k^{h}_{i} = 4$ and $k^{p}_{i} = 3$. Hence, $k^{h}_{i} = 4 \times M^{h}_{i}$, where $M^{h}_{i}$ is the total number of higher-order links incident on the node $i$. To provide an equal opportunity to the pair-wise and higher-order links for steering localization on a given node, we introduce the total number of pair-wise links $4$ times greater than the higher-order links, i.e., $M^{p} = 4 \times M^{h}$ for $k_{i}^{h} = k_{i}^{p}$.
Next, we define a parameter $\gamma = \frac{M^p}{4 \times M^{h}}$ to measure relative contribution for both the types of the links. Thus, if $\gamma > 1$ then $k_{i}^{p} > k_{i}^{h}$;  if $\gamma < 1$ then $k_{i}^{p} < k_{i}^{h}$ holds.

\begin{figure}[t]
\centering
\includegraphics[width= 0.48\textwidth]{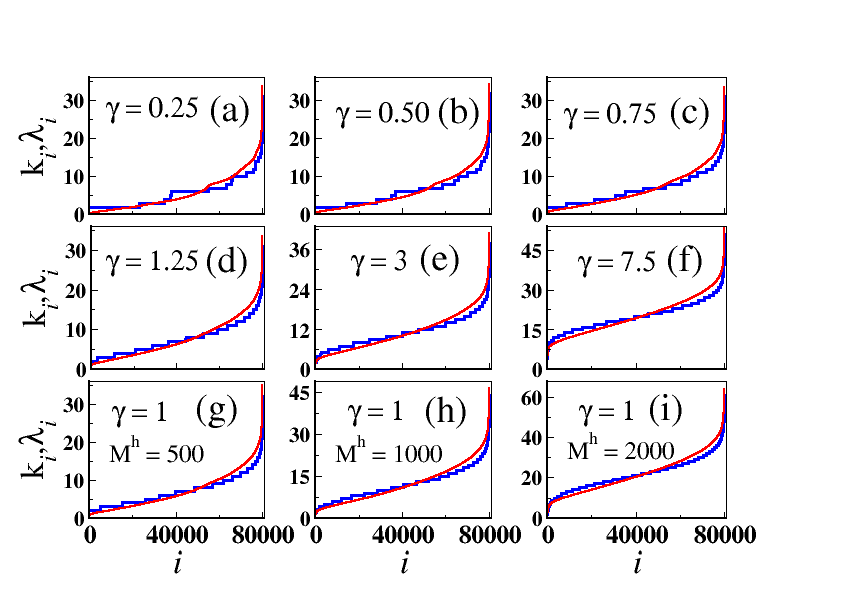} 
\caption{(Color online) Laplacian eigenvalues $\lambda_{i}$ (red) and node degree $k_{i}$ (blue) of hypergraph against index $i$ arranged in a increasing order for the $\gamma$ values. The size of the hypergraph $N= 2000$ and $M^{h} = 500$ remain fixed for all $\gamma$ values with $40$ random realizations.}
\label{deg_eig}
\end{figure}

We study the localization property of the eigenvectors of the hypergraph, and analyze the changes in the  localization behaviour as the pair-wise and the higher-order links are introduced on the initial ring lattice. Localization of an eigenvector  refers to a state that a few entries of the eigenvector have much higher values 
compared to the others. Degree of localization of the $\bm{x}_{j}^{th}$ eigenvectors can be quantified 
by measuring the inverse participation ratio (IPR) denoted as $Y_{\bm{x}_{j}}$. The IPR of an eigenvector $\bm{x}_{j}$ is defined as \cite{PEV_disease} 
\begin{equation}
Y_{\bm{x}_{j}}  = \sum_{i=1}^N (x_{i})_j^{4},  
\label{ipr}
\end{equation}
where $(x_{i})_j$ is the $i^{th}$ component of the normalized eigenvectors  $\bm{x}_{j}$ with $j$ $\in\left\{1,2,3 \ldots ,N\right\}$.\
One can easily verify that, for the most delocalized eigenvector all its components should be equal, i.e., $(x_{i})_j = \frac{1}{\sqrt{N}}$,  with IPR value being $1/N$. Whereas, for the most localized eigenvector, only one component of the eigenvector will be non-zero, and consequently IPR will be equal to $1$. 
It is also important to note here that there may exist fluctuations in the IPR  values for a given state $\bm{x}_{j}$ for different
realizations.  However, it is not possible that ${\lambda_{j}}$ remains the same for all the random realizations, and one has to be very careful in carrying the averaging. So, for the robustness of the results, we consider a small width $d \lambda$ around $\lambda$ and average all the IPR values corresponding to those $\lambda$ values which fall inside this small width \cite{eig_window, mishra_frac}. 

\begin{figure}[t]
\centering
\includegraphics[width= 0.48\textwidth]{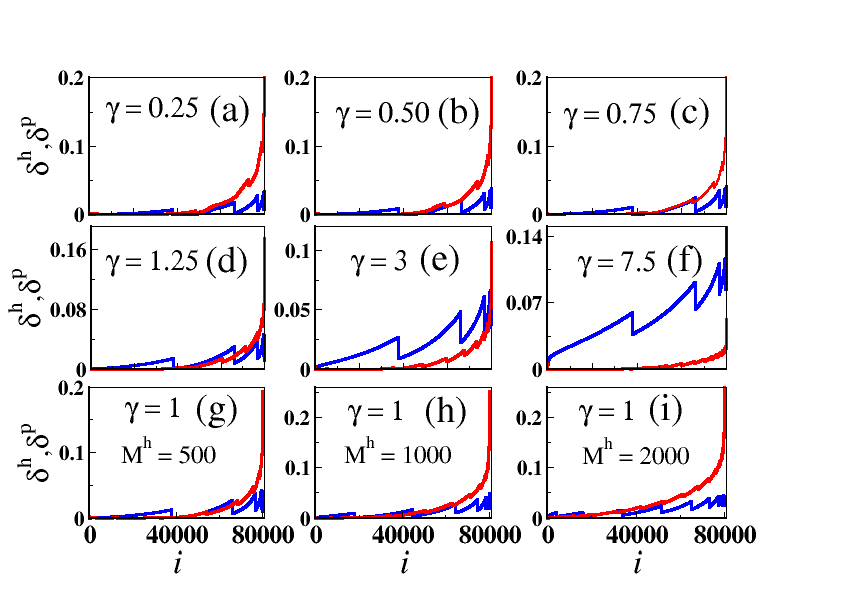} 
\caption{(Color online) The relative deviations of eigenvalues from the higher-order degrees,  $\delta_{i}^{h}$ (blue) and pair-wise degrees, $\delta_{i}^{p}$ (red) against index $i$. The hypergraph parameters, $N= 2000$ and $M^{h} = 500$, remain fixed for all $\gamma$ values with $40$ random realizations.}
\label{SWH_devi}
\end{figure}

We further elaborate on the averaging procedure for the discrete  eigenvalue spectrum achieved  through the numerical calculations as follow.\ Let $\lambda^{R}$ = \{$\lambda_{1}$,$\lambda_{2}$, $\ldots$ ,$\lambda_{N\times R}\}$ such that  
$\lambda_{1} \leq \lambda_{2} \leq $\ldots$ \leq \lambda_{N\times R}$  is a  set of eigenvalues of the hypergraph for all  $R$ random realizations where $N \times R$ is the size of $\lambda^{R}$.\ 
The corresponding eigenvector of $\lambda^{R}$ are denoted by {$\bm{x}^R$} = $\left\{\bm{x}_{1}, \bm{x}_{2}, \bm{x}_{3}, \ldots, \bm{x}_{N\times R}\right\}$.
We then divide $\lambda^{R}$ for a given value of $d\lambda$ into further $m$ subsets where $m$ = $(\lambda_{N \times R} - \lambda_{1}) $/$ d\lambda$.\  
For each
$\lambda^{j}$ $\subset$ $\lambda^{R}$ and  the corresponding eigenvectors $\bm{x}^{j}$ $\subset$ {$\bm{x}^R$},  $\forall j = 1,2,$\ldots$ ,m $; 
$\lambda^{j}$ = $\{\lambda_{1}, \lambda_{2},$\ldots$,\lambda_{l^{j}}\}$ and corresponding eigenvector $\bm{x}^{j}$ = $\left\{\bm{x}_{1}, \bm{x}_{2}, \bm{x}_{3}, \ldots, \bm{x}_{l^j}\right\}$  where $l^{j}$ is  the size of $j^{th}$  subset such
that $\sum_{j=1}^m l^{j}$ = $N \times R$ with a constraint that $\lambda_{l^{j}} - \lambda_{1^{j}} \leq d\lambda$. For each subset, the corresponding set of  IPRs for $\bm{x}^{j}$ 
will be $\{Y_{\bm{x}_{1}},Y_{\bm{x}_{2}}, $\ldots$, Y_{\bm{x}_{l^j}}\}$.\ 
Hence, the average IPR ($Y_{\bm{x}_{j}}(\lambda)$) for each subset $\bm{x}^{j}$ can be calculated as
$ \frac {\sum_{i=1}^{l^{j}} Y_{\bm{x}_{i}}} 
{l^{j}}$ where $\lambda$ is the central value for each subset, i.e., $\lambda$ + $\frac{d\lambda}{2}$ = $\lambda_{l^{j}}$ and $\lambda$ - $\frac{d\lambda}{2}$ = $\lambda_{1^{j}}$. Here, we define few more physical quantities, $k^{h}(\lambda)$, $k^{p}(\lambda)$, $\hat{k}^{h}(\lambda)$, $\hat{k}^{p}(\lambda)$  used in the paper. For any eigenvector $\bm{x}_{j}$, these quantities can be calculated as the following.  \\
$k^{h}_{\bm{x}_{j}}$ : higher-order degree of the node $i_{o}$ with the maximum component in $|(x_{i})_j|$,  i.e.,\ $(x_{io})_j$ $=$ max$\{|(x_{1})_j|, |(x_{2})_j|, \dots |(x_{N})_j|\}$ \\
$k^{p}_{\bm{x}_{j}}$ : pair-wise degree of the node $i_{o}$ with the maximum component in $|(x_{i})_j|$  i.e.\ $(x_{io})_j$ $=$ max$\{|(x_{1})_j|, |(x_{2})_j|, \dots |(x_{N})_j|\}$ \\
$\hat{k}^{h}_{\bm{x}_{j}}$ : higher-order degree expectation value of eigenvector, defined as  $\sum_{i=1}^N (x_{i})_j^{2} k^{h}_{i}$.\\
$\hat{k}^{p}_{\bm{x}_{j}}$ : pair-wise degree expectation value of eigenvector, defined as $\sum_{i=1}^N (x_{i})_j^{2} k^{p}_{i}$.\\

All these physical quantities follow the same averaging procedures over $\lambda$ and $\lambda + d\lambda$ as described for IPR and we obtain $k^{h}(\lambda)$, $k^{p}(\lambda)$, $\hat{k}^{h}(\lambda)$, $\hat{k}^{p}(\lambda)$.
\begin{figure}[ht]
\centering
\includegraphics[width= 0.48\textwidth]{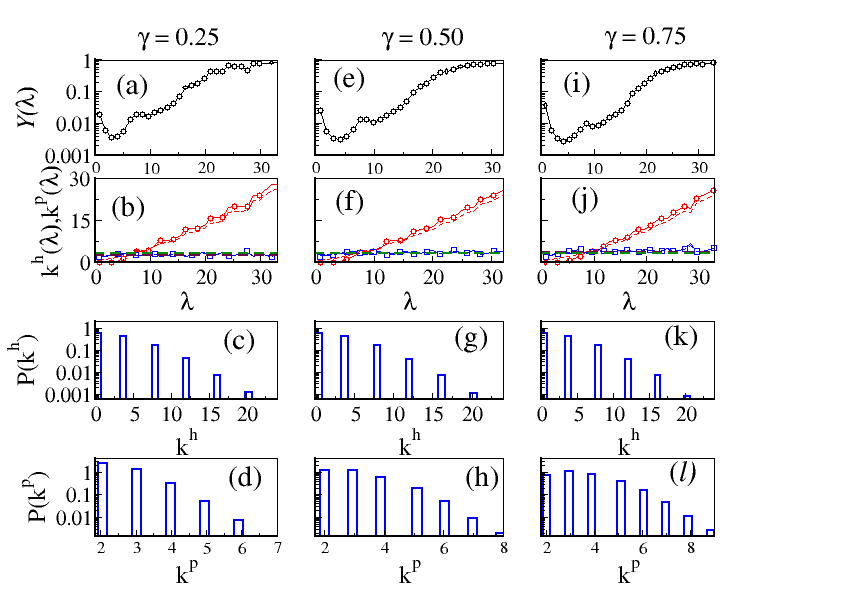} 
\caption{(Color online) Average IPR ($Y_{\bm{x}_{j}}(\lambda)$) ($\circ$),\  $k^{h}(\lambda)$ ($\color{red} \circ$ ), $k^{p}(\lambda)$ ($\color{blue}\square$), $\hat{k}^{h}(\lambda)$ ($\color{red} ---$), $\hat{k}^{p}(\lambda)$ ($\color{blue} ---$) against $\lambda$ for various $\gamma < 1$. The corresponding higher-order and pair-wise degree distribution are also plotted in last two rows. $\color{green} ---$ and $\color{brown} -\cdot-\cdot-$ are at $\langle k^{h} \rangle $ and $\langle k^{p} \rangle$ on y axis. The size of the hypergraph, $N= 2000$ and $M^{h} = 500$ remain fixed for all $\gamma$ values with $40$ random realizations.  }
\label{gam_0.9}
\end{figure}

\begin{figure}[ht]
\centering
\includegraphics[width= 0.48\textwidth]{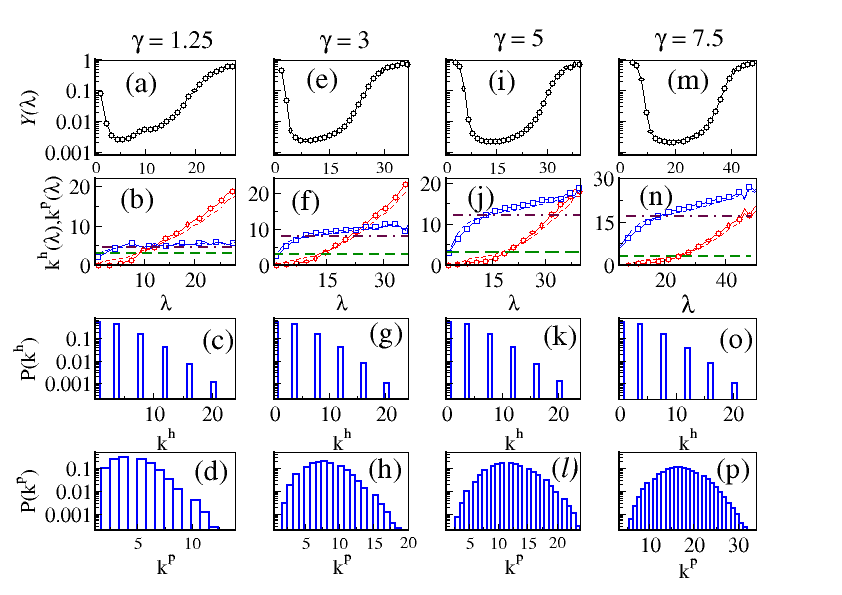} 
\caption{(Color online)  Average IPR ($Y_{\bm{x}_{j}}(\lambda)$) ($\circ$),\  $k^{h}(\lambda)$ ($\color{red} \circ$ ), $k^{p}(\lambda)$ ($\color{blue}\square$), $\hat{k}^{h}(\lambda)$ ($\color{red} ---$), $\hat{k}^{p}(\lambda)$ ($\color{blue} ---$) against $\lambda$ for various $\gamma > 1$. The corresponding higher-order and pair-wise degree distribution are also plotted in last two rows. $\color{green} ---$ and $\color{brown} -\cdot-\cdot-$ are at $\langle k^{h} \rangle $ and $\langle k^{p} \rangle$ on y axis. The size of the hypergraph, $N= 2000$ and $M^{h} = 500$ remain fixed for all $\gamma$ values with $40$ random realizations. }
\label{gam_1.1}
\end{figure}

\section{Results} \label{a4}
We first discuss the degree-eigenvalue correlation of the Laplacian of hypergraphs.
It was shown for pair-wise interactions \cite{eig_deg} that the eigenvalues of the Laplacian matrices have similar distributions as that of the degree of the nodes.
The relative average deviation between the eigenvalues and degrees of a network can be defined as $\frac{|| \lambda(L) - k ||_{2}}{||k||_{2}} \leq \sqrt{\frac{||k||_{1}}{||k||_{2}^{2}}}$, where $\lambda = (\lambda_{1}, \lambda_{2}, \dots \lambda_{N})^{T}$ and  $k = (k_{1},k_{2} \dots k_{N})^{T}$ are the  eigenvalue set of the Laplacian matrix and node degree set arranged in the increasing order, respectively. The $||y||_{p}$ represents the p-norms of any vector $y = (y_{1}, y_{2}, \dots y_{n})$ and is defined as  ($\sum_{i} {|y_{i}|}^{p})^{\frac{1}{p}}$. Thus, we see that $\sqrt{\frac{||k||_{1}}{||k||_{2}^{2}}} \ll 1$ which implies that eigenvalue distribution and degree distribution will have similar nature. Also, it is well known that $\langle k \rangle = \langle \lambda \rangle$. Fig.~\ref{deg_eig} plots the eigenvalues ($\lambda_{i}$) and degree ($k_{i}^{H}$) of the hypergraph arranged in increasing order with $N = 2000$ and $40$ random realizations for various $\gamma$ values. A clear degree-eigenvalue correlation can be seen similar to the pair-wise networks from the Fig.~ \ref{deg_eig}.

Next, it is difficult to decompose the eigenvalue $\lambda_{i}$ exactly into $\lambda_{i}^{h} + \lambda_{i}^{p} = f(k_{i}^{h})+f(k_{i}^{h})$, as done for the node degree. Nevertheless,  we can put few heuristic arguments as follows. First, we define $\delta_{i}^{h} = \frac{(|\lambda_{i}-k_{i}^{h}|)}{\lambda_{i}}$ and 
$\delta_{i}^{p} = \frac{(|\lambda_{i}-k_{i}^{p}|)}{\lambda_{i}}$ to analyze the relative deviations of the eigenvalues from the higher-order and pair-wise degrees. Fig.~\ref{SWH_devi}   plots  $\delta_{i}^{h}$,  $\delta_{i}^{p}$ against  $i$ for various $\gamma$ values. For  $\gamma \leq 1$, for  the initial eigenvalues ($\lambda \leq 6$) and index ($i \leq 40000$ ), $\delta_{i}^{h}$ $>$  $\delta_{i}^{p}$. Thus, the eigenvalues are more correlated with the pair-wise degree than that of the higher-order degree. Thus, $\lambda_{i}^{h} < \lambda_{i}^{p}$ and also $k_{i}^{h} = 0 $ for $i < 40000$ (not shown). For the intermediate eigenvalues, $\delta_{i}^{h}$ $\approx$  $\delta_{i}^{p}$ and thus $\lambda_{i}^{h} \approx \lambda_{i}^{p}$. For the extremal eigenvalues and large index, we see that 
$\delta_{i}^{h}$ $\ll$  $\delta_{i}^{p}$ and therefore  $\lambda_{i}^{p} \ll \lambda_{i}^{h}$. We  further mention here that for the said parameter, i.e., $\gamma < 1$, $M^{h} = 500$, $k^{p}_{max} = 10 $ and $k^{h}_{max} = 24$, and therefore for the extremal eigenvalues, the higher-order degrees play a governing role. For $1 <\gamma \leq 3$, $\delta_{i}^{h}$ $<$  $\delta_{i}^{p}$ for the extremal eigenvalues, since $k^{p}_{max} < k^{h}_{max}$, for $ 1 <\gamma \leq 3$. However, for $\gamma >3$, $\delta_{i}^{h}$ $>$ $\delta_{i}^{p} $ and thus $\lambda_{i}^{h} < \lambda_{i}^{p}$ for entire eigenvalue spectrum. Next, we discuss the interplay of higher-order and pair-wise links on instigating localization.
\\

\paragraph{{\bf For $\gamma <1$:}}  Figs.~\ref{gam_0.9} [a][e][i] plot results for $Y_{\bm{x}_{j}}(\lambda)$ against $\lambda$ for various values of $\gamma < 1$. The larger eigenvalues in the eigenvalue spectrum ($\lambda > 22 $) are localized with  $Y_{\bm{x}_{j}}(\lambda) \rightarrow 1$. 
For $\lambda \leq 22 $, eigenvalues are relatively less localized as compared with the larger eigenvalues. It is worth mentioning here that there exists no change in the nature of the plot with an increase in $\gamma$,  depicting the inefficacy of the pair-wise links on instigating localization.  Next, as reflected from  Figs.~\ref{gam_0.9} [b][f][j] that $k^{p}(\lambda)$ remains constant to a value around $\langle k^{p} \rangle$ for all the $\gamma$ values. Further, Ref.~ \cite{eig_window} argued that  localization of a eigenvector is centred on the nodes having very high or low degrees, or degree much deviating from the average degree of the corresponding network.  One can also note that for a completely delocalized eigenvector, degree expectation value $ \hat{k} = \sum_{i=1}^N (x_{i})_j^{2} k_{i} = $  
$\frac{(k_{1} + k_{2} + k_{3} \dots k_{N})}{N} = \langle k \rangle$. However, upon scrutinizing $k^{h}(\lambda)$ closely, we find that the its behavior is significantly different from  $k^{p}(\lambda)$. $k^{h}(\lambda)$ exhibits an increasing trend with the increase in the eigenvalue, and takes the maximum possible value in the localized region. Thus, the eigenvectors are localized on the set of the nodes with $k^{h}$ being  significantly higher than $\langle k^{h} \rangle$. Further, in the same plot, $k^{h}(\lambda)$ and $k^{p}(\lambda)$ are shown to be in good approximation with the $\hat{k}^{h}(\lambda)$ and $\hat{k}^{p}(\lambda)$ values, respectively.
Also, we plot the higher-order degree distribution $P(k^{h})$ and pair-wise degree distribution $P(k^{p})$ in Fig.~\ref{gam_0.9} [c][d][g][h][k][$\ell$]. Few interesting observations are; The number of nodes with $k^{h} > 16$, $k^{h} > 12$, and $k^{h} > 8$ are only $0.04 \%$, $0.64 \%$ and $4 \%$, respectively and still the eigenvectors are localized on these nodes in the localized region.  All these observations clearly suggest prime role of higher-order degree on instigating localization, and rather no visible impact of pair-wise links on the same.

\paragraph{{\bf For $\gamma >1$:}}  Here, we discuss the localization properties of the eigenvectors for $\gamma >1$. Figs.~\ref{gam_1.1} [a][e][i][m] present the results for $Y_{\bm{x}_{j}}(\lambda)$ as  a function of $\lambda$ for various values of $\gamma > 1$. The extremal part of the eigenvalue spectrum (larger and smaller eigenvalues) are highly localized with  $Y_{\bm{x}_{j}}(\lambda) \rightarrow 1$. On the contrary, the central part of the eigenvalue spectrum are delocalized with $ 10^{-3} \leq Y_{\bm{x}_{j}}(\lambda) < 10^{-2}$. 
Also, as $\gamma$ increases the eigenvectors corresponding to the smaller eigenvalues get more localized captured by the IPR value. 
The nature of the plots is similar to those of the small-world networks (pair-wise)  shown in \cite{eig_window}. Further, we report that the eigenvectors are localized on the nodes with degree ($k_{i}^{H}$) being abnormally high or low from the average degree ($\langle k \rangle$), which is  consistent with the observations made in \cite{eig_window,Locn_dyn}. 

Next, we discuss the role of pair-wise  ($k_{i}^{p}$) and higher-order links ($k_{i}^{h}$) separately on steering the localization.
Figs.~\ref{gam_1.1} [b][f][j][n] illustrate the results for $k^{h}(\lambda)$, $k^{p}(\lambda)$, $\hat{k}^{h}(\lambda)$, $\hat{k}^{p}(\lambda)$ for various $\gamma$
values. For $\gamma = 0.25$,  $k^{p}(\lambda)$ remains constant around $\langle k^{p} \rangle$, but deviates slightly at the extremal part of the eigenvalue spectrum.
On the contrary, $k^{h}(\lambda)$ manifests an increasing trend with the increase in the eigenvalues and takes large possible values in the localized region of the spectrum. As $\gamma$ increases, the pair-wise links also start playing role in driving the localization. First, the degree of localization of the eigenvectors corresponding to the smaller eigenvalues enhances. This can be explained as follows. As $\gamma$ increases, the number of pair-wise links also increases which in turn leads to an increase in $\langle k^{p} \rangle$. Thus, $k^{p}(\lambda)$ $\ll \langle k^{p} \rangle$ for the case of smaller eigenvalues, which consequently intensifies the degree of localization. It is important to note that, $\langle k^{h} \rangle = 3$ remains fixed for all the $\gamma$ values with the $M^{h} = 500$. Therefore, $k^{h}(\lambda)$ for smaller eigenvalues can not be much lesser than 
$\langle k^{h} \rangle$, therefore suggesting  no visible role of higher-order links in localization for smaller eigenvalues. It is worth mentioning here that a similar result was obtained for degree-eigenvalue correlation. 

\begin{table}
    \caption{Number of nodes common between the sets $N_{o}(k^{h})$ and $N_{o}(k^{p})$. For $N = 2000$ and $40$ random realizations.} 
    \begin{tabular}{|l|l|l|l|l|l}\hline
         & $\gamma = 3$ & $\gamma = 5$ & $\gamma = 7.5$ \\ \hline
      $N_{o}(k^{h} > 8)$ & 3227 & 3212  &     3193       \\ \hline
      $N_{o}(k^{h} > 12)$ & 585   & 582  &        611                 \\ \hline
      $N_{o}(k^{h} > 16)$ & 78   & 93  &   82                      \\ \hline
      $N_{o}(k^{h} > 20)$ &  10 & 7       &  13               \\ \hline
      $N_{o}(k^{p} > 8)$ & 31509 & 69698 & 70494            \\ \hline
      $N_{o}(k^{p} > 12)$ & 3418 & 582 &  79400           \\ \hline
      $N_{o}(k^{p} > 16)$ & 125 & 6698 & 428561            \\ \hline
      $N_{o}(k^{p} > 20)$ & 2 & 508 &    14487         \\ \hline
     $N_{o}(k^{h} > 8) \cap N_{o}(k^{p} > 8) $ & 1286 & 2806 & 3168 \\ \hline
     $N_{o}(k^{h} > 12) \cap N_{o}(k^{p} > 12) $ & 31 & 237 & 532\\ \hline
     $N_{o}(k^{h} > 16) \cap N_{o}(k^{p} > 16) $ & 0 & 7 & 46\\ \hline
     $N_{o}(k^{h} > 20) \cap N_{o}(k^{p} > 20) $ & 0 & 0  &1 \\ \hline
\end{tabular}
\label{tab:tab1}
\end{table}

\begin{figure}[b]
\centering
\includegraphics[width= 0.47\textwidth]{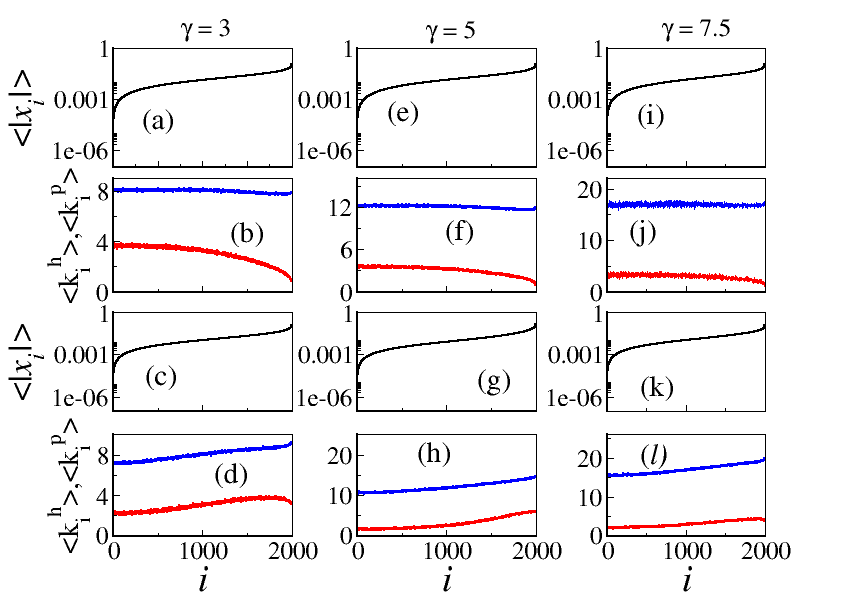} 
\caption{(Color online) $\langle |x_{i}| \rangle$ (black), $ \langle k_{i}^{h} \rangle$ (red) and $\langle k_{i}^{p} \rangle$ (blue) against index $i$ for $\lambda$ belonging to the delocalized region. (a)-(b) $\lambda \approx 8.5$, (c)-(d) $\lambda \approx 14 $, (e)-(f) $\lambda \approx 12.5$, (g)-(h) $\lambda \approx 23.5$, (i)-(j) $\lambda \approx 18$, (k)-($\ell$) $\lambda \approx 26$. The size of the hypergraph, $N= 2000$ and $M^{h} = 500$ remain fixed for all $\gamma$ values with $40$ random realizations. }
\label{SWH_dlocn}
\end{figure}

\begin{figure}[t]
\centering
\includegraphics[width= 0.47\textwidth]{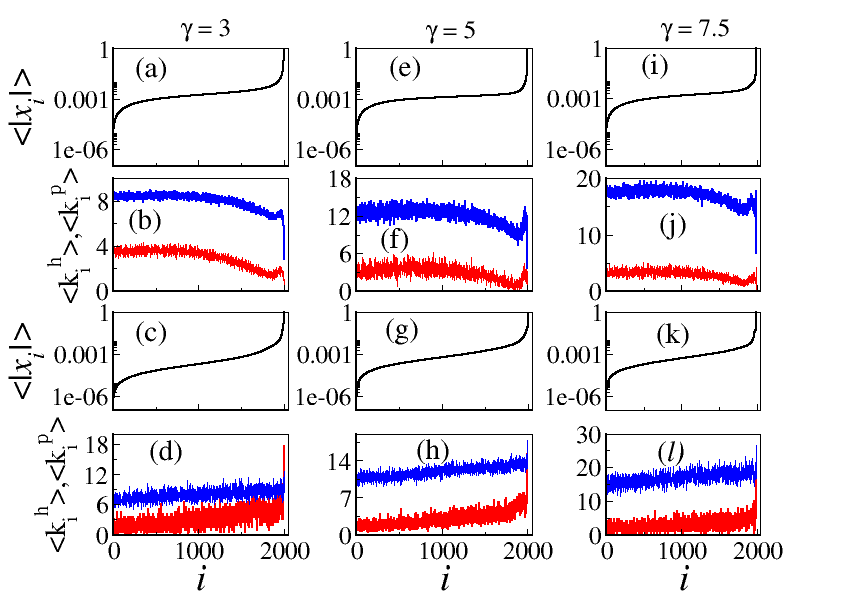} 
\caption{(Color online) $\langle |x_{i}| \rangle$ (black), $ \langle k_{i}^{h} \rangle$ (red) and $\langle k_{i}^{p} \rangle$ (blue) against index $i$ for $\lambda$ belonging to the localized region. (a)-(b) $\lambda \approx 2$, (c)-(d) $\lambda \approx 33 $, (e)-(f) $\lambda \approx 3$, (g)-(h) $\lambda \approx 36$, (i)-(j) $\lambda \approx 5$, (k)-($\ell$) $\lambda \approx 46$. The size of the hypergraph, $N= 2000$ and $M^{h} = 500$ remain fixed for all $\gamma$ values with $40$ random realizations. }
\label{SWH_locn}
\end{figure}

Next, we discuss the localization properties of the eigenvectors corresponding to those large eigenvalues which are highly localized. From the Fig.~\ref{gam_1.1}[b][f][j][n], it is visible that for $\gamma \geq 3$,  $k^{p}(\lambda)$ start deviating from $\langle k^{p} \rangle$, and hence, the pair-wise links also start participating in instigating localization along with the higher-order links. Few interesting things  to be noted here are; The number of nodes with large $k^{h}$ values are always very less as compared to the number of nodes with large $k^{p}$ values for $\gamma \geq 3$  (Fig.~\ref{gam_1.1} [c][g][k][o]). Scrutinizing more closely, we witness that the number of nodes with $k^{h} > 8, k^{h} > 12, k^{h} > 16$ and $k^{h} > 20$ are roughly around $4\% , 0.73 \% , 0.11 \% , 0.06 \%$, respectively for all $\gamma$ values. On the other hand. For $\gamma = 3$,
the number of nodes with $k^{p} > 8$ and $ k^{p} > 12$ are $39 \%$ and $4.27 \%$, respectively. For $\gamma = 5$, the number of nodes with $k^{p} > 12$ and $ k^{p} > 16$ are $41 \%$ and $8 \%$, respectively. For $\gamma = 7.5$, the number of nodes with $k^{p} > 16$ and $ k^{p} > 20$ are $53\%$ and $18 \%$, respectively.
Therefore, despite the number of nodes with large $k^{h}$ being very small, the higher-order links still keep playing very crucial roles in steering localization for the larger eigenvalues.
To get further insight  into the role of higher-order links, we define the following quantities. Let $N_{o}(k^{h} > c)$ and  $N_{o}(k^{p} > c)$ denote the set of nodes with 
$k^{h} > c$ and $k^{p} > c$, respectively. We are interested to find out the number of nodes which are common between these two sets, i.e., $N_{o}(k^{h} > c) \cap N_{o}(k^{p} > c) $
 (Table~\ref{tab:tab1}). It is evident from the table that total number of nodes in $N_{o}(k^{h} > c) \cap N_{o}(k^{p} > c) $ is less than $50 \%$ of the set $N_{o}(k^{h} > c)$ for all $\gamma$ values and $c>8$. Therefore, the impact of the higher-order links, over the pair-wise links, in steering localization for larger eigenvalues is apparent more profoundly, which can be attributed to the increase in $\langle k^{p} \rangle$ value with the increase in $\gamma$, while $\langle k^{h} \rangle = 3$ taking a constant value. Thus, $k^{h}(\lambda) \gg \langle k^{h} \rangle$ for the case of large eigenvalues for all the $\gamma$ values. On the contrary, though the largest pair-wise degree $k^{p}_{max}$ experience an increase with the increase in $\gamma$, at the same-time $\langle k^{p} \rangle$ also increases, and hence  $k^{h}(\lambda) - \langle k^{h} \rangle$ $>$ $k^{p}(\lambda) - \langle k^{p} \rangle$. We obtain the same results are obtained for $\gamma \leq 15$ (not shown).

So far, we have discussed the localization properties of the eigenvectors by using  $k^{h}(\lambda)$ and  $k^{p}(\lambda)$, and demonstrated that eigenvectors are localized on the node having the degree abnormally high or low either with respect to $\langle k^{h} \rangle$ or $\langle k^{p} \rangle$. Also, $k^{h}(\lambda)$ and  $k^{p}(\lambda)$ come in the good approximation with $\hat{k}^{h}(\lambda)$ and $\hat{k}^{p}(\lambda)$. However, it is also important to scrutinize other eigenvector components to obtain
the holistic idea of the localization. For this, we calculate the absolute value of the eigenvector components $|x_{i}|$, the higher-order degree of the corresponding node $k_{i}^{h}$, the pair-wise degree $k_{i}^{p}$, and  average them over $\lambda$ and $\lambda \pm d\lambda$ denoted by $\langle |x_{i}| \rangle$, $ \langle k_{i}^{h} \rangle$ and $\langle k_{i}^{p} \rangle$. We consider different regions of the eigenvalue spectrum to calculate  $\langle |x_{i}| \rangle$, $ \langle k_{i}^{h} \rangle$ and $\langle k_{i}^{p} \rangle$. For the delocalized region, we consider the $\lambda$ values where $k^{h}(\lambda)$ intersects with $\hat{k}^{h}(\lambda)$, and $k^{p}(\lambda)$ intersects with $\hat{k}^{h}(\lambda)$. For the localized region, we take $\lambda$ from the extremal eigenvalues, i.e., smaller and larger eigenvalues. 
Fig.~\ref{SWH_dlocn} presents the results for $\langle |x_{i}| \rangle$ arranged in an increasing order, and corresponding $ \langle k_{i}^{h} \rangle$ and $\langle k_{i}^{p} \rangle$ for two $\lambda$ values belonging to the delocalized region for various $\gamma$ values. It is clearly visible that max($\langle |x_{i}| \rangle$) $\ll 1$  and most of the $\langle |x_{i}| \rangle$ are in the order of $10^{-2}$ and $10^{-3}$, respectively. It becomes more interesting to look at the behavior of $ \langle k_{i}^{h} \rangle$ and $\langle k_{i}^{p} \rangle$. Both $\langle k_{i}^{h} \rangle$ and $\langle k_{i}^{p} \rangle$ remain constant to values which are around $\langle k^{h} \rangle $ and $\langle k^{p} \rangle$, respectively and thus validating the earlier results.
Further, in Fig.~\ref{SWH_locn} plots $\langle |x_{i}| \rangle$, $ \langle k_{i}^{h} \rangle$ and $\langle k_{i}^{p} \rangle$ for $\lambda$ belonging to the localized region. It is apparent from that max($\langle |x_{i}| \rangle) \rightarrow 1$,  and only a few entries are in the order of $10^{-1}$ depicting the localized nature of the eigenvectors. Further, for smaller eigenvalues, $\langle k_{i}^{h} \rangle$ remain fixed to the values which lie in the close vicinity to $\langle k^{h} \rangle \approx 3\pm 2$ for all $i$, however, $\langle k_{i}^{p} \rangle$ deviates from $\langle k^{p} \rangle$ and dips down to a value which is lower than $\langle k^{p} \rangle$ for the node contributing maximum in 
$\langle |x_{i}| \rangle$. For the larger eigenvalues, $\langle k_{i}^{h} \rangle$  remains constant at around $\langle k^{h} \rangle$ for the nodes contributing minimal in  
$\langle |x_{i}| \rangle$, and takes value much larger than $\langle k^{h} \rangle$ for the nodes contributing maximal in $\langle |x_{i}| \rangle$. On the other hand, $\langle k_{i}^{p} \rangle$ always keeps oscillating around $\langle k^{p} \rangle$ for all $i$, and shows a little deviation for $\langle k^{p} \rangle$ for large $i$.
The above observations clearly validate the earlier obtained result that localization at smaller eigenvalues is instigated by the pair-wise links, with higher-order links playing a dominant role in inducing localization for larger eigenvalues.

\section{Conclusion} \label{a5}
To conclude, we have investigated an interplay of the higher-order and the pair-wise links in instigating the localization of the eigenvectors of the hypergraphs. We find that eigenvectors are localized on the set of nodes having degrees either  much higher or lower from the average degree, a result which is consistent with the earlier known result for the networks having only pair-wise interactions. Further, by defining a single parameter $\gamma$ on each node, we show that there exists no impact of pair-wise links on localization for $\gamma \leq 1$.  
For $\gamma >1$, with an increase in $\gamma$, the degree of the localization of the eigenvectors corresponding to the smaller eigenvalues increases. Also, the role of higher-order links is not significant as compared to the pair-wise links in inducing localization for smaller eigenvalues. This is due to the fact that $\langle k^{p} \rangle$ increases with the increase in $\gamma$, but $\langle k^{h} \rangle = 3$ remains to a fixed value.
Thus, $k^{h} (\lambda)$ being small for lower eigenvalues can not be substantially low as compared with  $\langle k^{h} \rangle$ for all $\gamma$ values.
On the contrary, the difference between the pair-wise degree of a node contributing maximum in the eigenvector ($k^{p} (\lambda)$) and $\langle k^{p} \rangle$ starts increasing with the increase in $\gamma$. Ergo, the pair-wise links play a significant role in the eigenvector localization for smaller eigenvalues. Whereas, for larger eigenvalues, the higher-order links play a crucial role in instigating localization despite the fact that the number of nodes with high value of higher-order degree ($k^{h}$) remains very small for all the $\gamma$ values. This can also be explained in a similar fashion which we adopted for the smaller eigenvalues. As $\langle k^{h} \rangle = 3$ remains fixed to a constant value for all $\gamma$ values, the difference between $k^{h} (\lambda)$ and $\langle k^{h} \rangle$ for larger eigenvalues always remain very high, while for the pair-wise links though the largest pair-wise degree $k^{p}_{max}$ exhibits an increase with $\gamma$, there exists a simultaneous increase in $\langle k^{p} \rangle$. Therefore, $k^{h}(\lambda) - \langle k^{h} \rangle$ $>$ $k^{p}(\lambda) - \langle k^{p} \rangle$ for larger eigenvalues, which in turn indicates importance of  higher-order links on the localization for larger eigenvalues.
The present work can be extended to simplicial complexes in which  higher-order degrees can be further decomposed into contributions
coming from different dimensions such as triangles, tetrahedrons, and so on. Further,  investigating roles of higher-order and pair-wise interactions, separately,  on dynamics of random walkers is an interesting future direction.

\section {Acknowledgements}
 S.J. acknowledges Government of India, Department of Science and Technology grant SPF/2021/000136.

\providecommand{\noopsort}[1]{}\providecommand{\singleletter}[1]{#1}%
%


\end{document}